\documentstyle[prl,aps,epsf]{revtex}\def\narrowtext{}\tighten\twocolumn
\input epsf.sty
\begin{document}
\draft

\title{
Ni-substituted sites and the effect on Cu electron spin dynamics of YBa$_2$Cu$_{3-x}$Ni$_x$O$_{7- \delta}$
}
\author{Y. Itoh$^{1,2}$, S. Adachi$^1$, T. Machi$^1$, Y. Ohashi$^3$, and N. Koshizuka$^1$}

\address{
$^1$Superconductivity Research Laboratory,
 International Superconductivity Technology Center,\\
1-10-13 Shinonome, Koto-ku Tokyo 135-0062, Japan \\
$^2$Japan Science and Technology Corporation, 4-1-8 Honcho,
Kawaguchi, Saitama 332-0012, Japan\\
$^3$Institute of Physics, University of Tsukuba, Ibaraki 305-8571, Japan\\
}

\date{\today}%
\maketitle %

\begin{abstract}%
We report Cu nuclear quadrupole resonance experiment on magnetic impurity 
Ni-substituted
YBa$_2$Cu$_{3-x}$Ni$_x$O$_{7- \delta}$. The distribution of Ni-substituted 
sites and its effect on the Cu electron spin dynamics are investigated. 
Two samples with the same Ni concentration $x$=0.10 and nearly the same oxygen content 
but different $T_c$'s were prepared:
One is an as-synthesized sample (7-$\delta$=6.93) in air ($T_c$$\approx 80 K$), and the other is a quenched one (7-$\delta$=6.92) in a
reduced oxygen atmosphere ($T_c$$\approx 70 K$). The plane-site $^{63}$Cu(2) nuclear spin-lattice relaxation 
for the quenched sample was faster
than that for the as-synthesized sample, in contrast to the $^{63}$Cu(1) relaxation that was faster for the as-synthesized sample.
This indicates that the density of plane-site Ni(2) is higher in the quenched sample, 
contrary to the chain-site Ni(1) density which is lower in the quenched sample. 
From the analysis in terms of the Ni-induced nuclear spin-lattice
relaxation, we suggest that the primary origin of suppression of $T_c$ is associated with nonmagnetic depairing effect 
of the plane-site Ni(2). 
\end{abstract}

\narrowtext

Magnetic impurity causes depairing in both $s$-wave and $d$-wave superconductivity.
Complete suppression of the superconducting transition
temperature $T_c$ is observed for Ni-doped La$_{2-x}$Sr$_x$CuO$_4$ 
and YBa$_2$Cu$_4$O$_8$, somewhat more weakly
than for Zn-doped ones~\cite{Xiao,Watanabe,Miyatake}. 
Ni impurity in the high-$T_c$ cuprate superconductor
carries a localized moment, because the uniform spin susceptibility with Curie or Curie-Weiss law
is observed. The depairing effect of potential scatterer Zn on the $d$-wave superconductivity
is a natural consequence from breakdown of Anderson's theorem~\cite{Millis,Balian}. 
For YBa$_2$Cu$_3$O$_{7-\delta}$ with the optimized $T_c$=92 K, however, the decrease of $T_c$ per Ni concentration is
smaller than that per Zn concentration~\cite{Markert,TM,Tarascon}. 
Figure 1(a) shows the impurity doping dependences of $T_c$ for various high-$T_c$ superconductors
with Ni or Zn~\cite{Xiao,Tokunaga1,Kluge,Itoh,Itoh1}.
The decrease of $T_c$ by Ni doping for YBa$_2$Cu$_3$O$_{7-\delta}$ is smallest among these compounds with Ni in Fig. 1(a).
This small decrease of $T_c$ by Ni doping for YBa$_2$Cu$_3$O$_{7-\delta}$ was attributed 
to a weak scattering of conducting carriers by Ni
impurities (Born scatterer)~\cite{Ishida} or to a softening of pairing frequency itself~\cite{Tokunaga1}.  
Nevertheless, it has been suspected that only a part of the doped Ni impurities is substituted for the plane Cu(2) site 
and that the remaining part is for the chain Cu(1) site~\cite{Howland,Yang,Bridges,Morita} 
(see the references in Ref.~\cite{Adachi}). 
The bulk $T_c$ is considered to be determined by the amount of the in-plane Ni(2) impurity.  
One should note that YBa$_2$Cu$_3$O$_{7- \delta}$ has two crystallographic Cu sites, i.e.
 the chain Cu(1) and the plane Cu(2) sites.
The recent observation of an in-plane anisotropy of optical
conductivity for detwinned single crystal YBa$_2$Cu$_{3-x}$Ni$_x$O$_{7-\delta}$ 
indicates the existence of the chain-site Ni(1)~\cite{Homes}. 
A microscopic study on the selective substitution of Ni impurity is therefore of interest. 
 
Synthesis of oxides under reduced oxygen partial pressure is frequently 
effective in controlling cation solid solution, e.g. to synthesize the superconducting 
La$_{1+x}$Ba$_{2-x}$Cu$_3$O$_{7-\delta}$~\cite{Wada}, or to optimize the 
superconducting critical current density or the irreversible magnetic field of 
Nd$_{1+x}$Ba$_{2-x}$Cu$_3$O$_{7-\delta}$~\cite{Yoo,Shibata}. Recently, using the 
reduced oxygen partial pressure technique, Adachi et al., succeeded in controlling 
$T_c$ of YBa$_2$Cu$_{3-x}$Ni$_x$O$_{7-\delta}$ with optimal oxygen content 
~\cite{Adachi}. They synthesized two series of YBa$_2$Cu$_{3-x}$Ni$_x$O$_{7-\delta}$ 
samples having the different $T_c$'s per Ni concentration. 
Figure 1(b) shows the Ni doping dependence of $T_c$ of their samples~\cite{Adachi}. 
Here, we focus on two samples with the same Ni concentration $x$=0.10 ($z$=0.033 in Fig. 1(b)) and nearly the same oxygen
content but different $T_c$. 
For convenience, let us call one sample with $T_c$$\approx 80 K$ 
(7-$\delta$=6.93) an as-synthesized one, because it was synthesized in flowing oxygen gas without quenching treatment, and the other
sample  with $T_c$$\approx 70 K$ (7-$\delta$=6.92) a quenched one,
because it was the as-synthesized sample once again fired and quenched in a reduced 
oxygen atmosphere at 800 $^\circ$C. The details are given in Ref.~\cite{Adachi}. 
Ni prefers the higher coordination of oxygen atoms~\cite{Navrotsky,TM}. The 
plane-site Cu(2) is located in the pyramid with five oxygen ions, whereas the chain-site 
Cu(1) is coordinated with two, three, or four nearest neighbor oxygen ions. 
The two series of samples with different $T_c$ 
suggest that the distribution of Ni-substituted sites over Cu(1) and Cu(2) sites 
is changed through synthesis under the reduced oxygen atmosphere. 

In this paper, we report the measurements of Cu(1) and Cu(2) 
nuclear quadrupole resonance (NQR) spectra and nuclear 
spin-lattice relaxation curves, to study microscopically
the distribution of Ni-substituted sites for the above mentioned two samples 
(as-synthesized and quenched ones). The observed difference in the $^{63}$Cu nuclear spin-lattice relaxation at Cu(1) and Cu(2)
in the two samples must be a Ni doping effect. 
From the Cu NQR measurements, we give proof that the heat treatment
in reduced oxygen atmosphere results in a redistribution of the Ni atoms in YBa$_2$Cu$_{3-x}$Ni$_x$O$_{7-\delta}$.

	Two samples with precisely the same mole number could not be prepared, 
because some parts of the samples have already been used for the characterization~\cite{Adachi}. 
Hence, quantitative comparison of the relative intensity of Cu NQR spectra
could not be made to estimate the relative number of the observed nuclei.
The nuclear spin-lattice relaxation is independent of the sample volume
and is relatively more sensitive to the impurity than the intensity of NQR spectrum.
The powder samples were coated in paraffin oil. 
A coherent-type pulsed spectrometer was utilized for the zero-field Cu NQR 
measurements. The Cu NQR frequency spectra with quadrature detection were 
measured by integration of the spin-echoes as a function of rf frequency. 
The Cu nuclear spin-lattice relaxation curves were measured by
an inversion recovery spin-echo method, where the Cu nuclear spin-echo amplitude $M(t)$ was 
recorded as a function of time interval $t$ after an inversion $\pi$ pulse, 
in a $\pi-t-\pi/2-\pi$-echo sequence, and $M(\infty)$ was also recorded in a $\pi/2-\pi$-echo sequence
(no inversion pulse) as usual~\cite{Itoh,Itoh1,Itoh2,Itoh3}.  

Figure 2 shows the Cu NQR spectra at $T$=4.2 K. The amount of the as-synthesized sample is more than
that of the quenched one, nevertheless the intensity of
Cu(1) NQR spectra for the as-synthesized sample was weaker than that for the quenched 
one. But, we could not find qualitative difference in the line profiles of Cu NQR 
spectra between two samples.   

Figure 3 shows the $^{63}$Cu nuclear spin-echo recovery curves (spin-lattice 
relaxation curves)  
$p(t)\equiv1-M(t)/M(\infty)$ for Ni-doped samples at $T$=4.2 K. Inset figures 
show the recovery curves for pure (impurity-free) YBa$_2$Cu$_3$O$_{6.98}$ 
($T_c$=92 K). Solid curves are the least-squares fitting results using the theoretical function 
described below. 
First, from comparison with inset figures, all the recovery curves 
for both Ni-doped samples recover more quickly than those for pure 
YBa$_2$Cu$_3$O$_{6.98}$. Thus, Ni impurities distribute over both sites of Cu(1) 
and Cu(2). 
Second, the Cu(2) nuclear spin-echo signal 
recovers faster in the quenched sample than in the as-synthesized one, whereas the Cu(1) nuclear 
spin-echo signal recovers slower in the former than in the latter.
The Cu(2) nuclear spin-echo signal is affected in the quenched 
sample more than in the as-synthesized one, whereas the Cu(1) is vice versa. 
Thus, it is natural to conclude that the amount of Ni(2) in the quenched sample is more than that 
in the as-synthesized one and that of N(1) is vice versa.  

The $^{63}$Cu nuclear spin-echo recovery curves for Ni-doped samples in 
Fig. 3 are nonexponential functions. For quantitative discussion, we analyzed the 
experimental recovery curve $p(t)$ by the exponential function times a stretched 
exponential function  $p(t)=p(0)$exp$[-wt/(T_1)_{HOST}-\sqrt{wt/\tau_1}]$ 
($p(0)$, $(T_1)_{HOST}$ and $\tau_1$ are the fit parameters, and $w$=3 at Cu(2) and $w$=1 at
Cu(1)), after the dilute magnetic alloy~\cite{McHenry} or YBa$_2$Cu$_4$O$_8$ 
with Ni impurities~\cite{Itoh2}. 
The multiplicative numerical factor $w$ is introduced to conform to the conventional expression of $T_1$~\cite{Moriya0},
and it is not essential in the below discussion.
Here, $w=3$ is defined for the Cu(2) NQR $T_1$ under a uniaxial electric field gradient.   
The Cu(1) site is under an asymmetric electric field gradient~\cite{Pennington,Shimizu}, so that all the $x$-, $y$-, and
$z$-components of the fluctuating local field contribute to the Cu(1) NQR $T_1$. Then, we use a simple $w$=1.  
$(T_1)_{HOST}$ is the Cu nuclear spin-lattice  relaxation time due to the host Cu electron spin fluctuation via a hyperfine
coupling. 
$\tau_1$ is the impurity-induced nuclear spin-lattice relaxation time via 
a longitudinal direct dipole coupling or a two dimensional Ruderman-Kittel-Kasuya-Yosida 
interaction~\cite{Bobroff}. We have confirmed no significant
contribution from nuclear  spin diffusion by measuring $^{65}$Cu isotope dependence and 
pulse-strength $H_1$ dependence of the recovery curves~\cite{Khutsishvili}.
Although for the heavily impurity-doped system~\cite{Williams} or spin glass system~\cite{Kukovitsky} where it is hard to
assign separately the host and the guest contributions, a single stretched exponential function 
$p(t)=p(0)$exp$[-(t/\tau_1)^\alpha]$ with a variable exponent $\alpha$
might be appropriate,
we believe that the present model with two time constants, $(T_1)_{HOST}$ and $\tau_1$, is minimal
and appropriate for the $x$=0.10 samples. 

Figure 4 shows the temperature dependence of the estimated 
$(1/T_1)_{HOST}$ (a) [(c)] and of 1/$\tau_1$ (b) [(d)] at $^{63}$Cu(2) 
[$^{63}$Cu(1)]. Below $T_c$, the Ni-induced relaxation component (stretched exponential part)
is predominant both at Cu(1) and Cu(2). 
In Figs. 4(b) and 4(d), the Ni-induced relaxation rate 1/$\tau_1$ of Cu(2) for the quenched 
sample is more enhanced than that for the as-synthesized one, whereas that of Cu(1) is 
vice versa. In Fig. 4(b), the upturn of 1/$\tau_1$ of Cu(2) below 10 K for 
the as-synthesized sample is consistent with the upturn of an initial relaxation rate 
1/$T_{1s}$ reported in Ref.~\cite{Tokunaga2}. 
In Fig. 4(d), the difference in 1/$\tau_1$ of Cu(1) between the two samples with
Ni is larger at lower temperatures than about 30 K. It is hard to estimate precisely the small 1/$\tau_1$
and its difference above about 30 K, where the signal-to-noise ratio of Cu(1) NQR
is poor in the Ni-doped samples.

In Fig. 4(a), the host relaxation rate $(1/T_1)_{HOST}$ of Cu(2) in the as-synthesized and the quenched samples
decreases more steeply than that in pure YBa$_2$Cu$_3$O$_{6.98}$, as the temperature is decreased 
below $T$=18$\sim$20 K and below $T$=25$\sim$30 K, respectively. 
The dashed lines are $T^3$ functions characteristic of $d$-wave superconductivity. 
The steep decrease of $(1/T_1)_{HOST}$ and the upturn of $1/\tau_1$ is also observed for 
YBa$_2$Cu$_{4-x}$Ni$_x$O$_8$ with $x$=0.12 ($T_c$=15 K)~\cite{Itoh2}. 
It is theoretically suggested that
the spin-orbit coupling between an itinerant electron and a Ni local moment 
induces a local superconducting state with a different order parameter ($d_{xy}$-wave symmetry) 
around Ni in the $d_{x^2-y^2}$-wave  superconducting state~\cite{Balatsky}. 
The steep decrease of $(1/T_1)_{HOST}$ far below $T_c$ may indicate a Ni-induced impurity band associated 
with the different order parameter, more gapped on the Fermi surface than a pure $d_{x^2-y^2}$-wave gap.

Figure 5 shows the temperature dependence of $(1/T_1)_{HOST}$ (a) and 
$1/\tau_1$ (b) at Cu(2) in linear scale to show up the above $T_c$ data. 
The host $(1/T_1)_{HOST}$ of Cu(2) above $T_c$ is nearly independent of Ni doping, 
whereas $1/\tau_1$ increases with decreasing $T_c$.
In general, 1/$\tau_1$ is an increasing function of the impurity concentration~\cite{McHenry}.
Thus, the observed increase of 1/$\tau_1$ indicates the systematic increase of the in-plane Ni(2) concentration
$x_{plane}$. 

In the theoretical model of superconducting pairing mediated by antiferromagnetic spin fluctuations, 
the external depairing effect on $T_c$ is written by $T_c=T_{c0}-\Delta T_c$ with $T_{c0}\sim\Gamma_0(Q)\chi_0(Q)$
($\Gamma_0(Q)$ is the host  antiferromagnetic spin-fluctuation frequency,
and $\chi_0(Q)$ is the static staggered spin susceptibility)~\cite{Moriya,Monthoux} and with
$\Delta T_c\propto x_{plane}$~\cite{AG}. 
In the spin-fluctuation theory~\cite{Moriya1}, one can obtain
$(1/T_1)_{HOST}\propto TC/(T+\Theta)$ ($C\propto\chi_0(Q)/\Gamma_0(Q)$, and $\Theta\propto\Gamma_0(Q)$~\cite{Itoh6}) 
in the leading order. The actual fit result by this function is the dotted curve in Fig. 5(a).  
Thus, the quantitative temperature dependence of $(1/T_1)_{HOST}$  
tells us the characteristic spin-fluctuation parameters, $\chi_0(Q)$ and $\Gamma_0(Q)$, which may describe the pairing
interactions. 
The nearly Ni-independent $(1/T_1)_{HOST}$ in Fig. 5 indicates that the host
spin-fluctuation spectrum is nearly invariant under Ni doping.
In Ref.~\cite{Tokunaga1}, the host
$(1/T_1)_{HOST}$ was estimated to be systematically enhanced by Ni doping, which was regarded as evidence for softening
of the spin-fluctuation frequency
$\Gamma_0(Q)$ and for the central origin to reduce $T_c$. 
However, our analysis indicates that the Ni-enhanced Cu(2) nuclear 
spin-lattice relaxation comes from the extra relaxation in the stretched exponential part. 
The central origin to reduce $T_c$ is the external pair 
braking effect due to the increase of the in-plane Ni(2) concentration $x_{plane}$ in $\Delta T_c$. 
This is consistent with the original result for YBa$_2$Cu$_{4-x}$Ni$_x$O$_8$.
However, it is still hard to estimate the quantitative value of $x_{plane}$.  
 
Here, we analyze the temperature dependence of 1/$\tau_1$ and discuss the pair breaking mechanism of Ni. 
If an impurity spin-spin relaxation process dominates the impurity spin correlation, 1/$\tau_1$ may not
change with temperature~\cite{McHenry}. The widely known expression of 1/$T_1$ or 1/$\tau_1 \propto
(A/\hbar)^2S(S+1)/\omega_{e}$ ($A$ is the coupling constant of the nuclear moment to an electron, $S$ is the
electron spin, $\omega_{e}$ is the exchange frequency of mutual spin interaction) is derived from exchange narrowing
limit~\cite{Moriya0,Anderson}.  
However, the estimated 1/$\tau_1$ changes with temperature as in Figs. 4(b) and 4(d).
This indicates a significant contribution from an impurity spin-lattice relaxation process. 

The temperature dependences of 1/$\tau_1$ far below $T_c$ in Figs. 4(b) and 4(d) are different 
between two Ni-doped samples. Both the impurity spin-lattice relaxation
rate $\Gamma_{SL}$ and the impurity spin-spin relaxation rate $\Gamma_{SS}$ 
contribute to the actual total impurity relaxation rate $\Gamma_L$, e.g.
$\Gamma_L=\Gamma_{SL}+\Gamma_{SS}$.  As temperature is decreased, $\Gamma_{SL}$ decreases moderately but rapidly below
$T_c$~\cite{Mali,Imai}, and then $\Gamma_{SS}$ can play an important role in $\Gamma_L$.
In general, $\Gamma_{SS}$ is proportional to the number density of impurity $x_{imp}$ but $\Gamma_{SL}$ is independent of $x_{imp}$,
and 1/$\tau_1$ is proportional to $x_{imp}^2S(\omega_N/\Gamma_L)$ ($S(\omega_N/\Gamma_L)$ is the impurity magnetic
spectral function at the nuclear resonance frequency $\omega_N/2\pi$)~\cite{McHenry,Moriya0}. 
Thus, the difference in 1/$\tau_1$ may result from the different weight of
$\Gamma_{SS}$ with $x_{imp}$. In the below analysis, however, we neglect $\Gamma_{SS}$ 
and then the difference in the temperature dependence of 1/$\tau_1$ for simplicity.  

For an isolated local moment system in a conventional metal,
the dynamical spin susceptibility as a function of frequency $\omega/2\pi$ is expressed by $\chi(q,\omega) =
\chi_L/(1-i\omega/\Gamma_L)$, where $\chi_L\propto S(S+1)/T$ is the static spin susceptibility and
$\Gamma_L=\alpha T\equiv 4\pi$($JN_F)^2k_BT/\hbar$ ($J$ is the coupling constant of 
the localized moments to the band, and $N_F$ is the density of states at the Fermi level per spin) 
is the impurity fluctuation frequency due to Korringa relaxation~\cite{Hewson}.  
Then, one obtains 1/$T_1$ or 1/$\tau_1\propto
\Gamma_L/(\omega_N^2+\Gamma_L^2)$~\cite{BPP,Moriya0}. 
Since 1/$\tau_1\propto 1/\alpha T$ at high temperatures ($T \gg \omega_N/\alpha$) and
1/$\tau_1\propto \alpha T/\omega_N^2$ at low temperatures ($T\ll\omega_N/\alpha$), 
then 1/$\tau_1$ takes a maximum value at $T=\omega_N/\alpha$ ($\Gamma_L=\omega_N$).

In Figs. 4(b) and 4(d), 1/$\tau_1$ of $^{63}$Cu(2) and of $^{63}$Cu(1) takes a maximum 
at about 60 K and 10$\sim$20 K, respectively, 
which can be associated with the maximum at $T=\omega_N/\alpha$.
Assuming the Korringa relaxation $\Gamma_L$=4$\pi$($JN_F$)$^2k_BT$/$\hbar$ 
and putting $\Gamma_L$=$\omega_N$=22 MHz for
Cu(1) at $T$=10 K and $\Gamma_L$=31.5 MHz for Cu(2) at $T$=60 K, 
one obtains $\left|JN_F\right|$=0.5$\sqrt{\hbar \omega_N /\pi k_BT}=$0.003 for Cu(1) and $=$0.0014 for Cu(2). 
From a typical value of $N_F$=1.5 states/eV-spin direction, $\left|J\right|$ is 1.9 meV for Cu(1) and 0.94 meV for Cu(2).
The actual maximum of 1/$\tau_1$ takes place in the superconducting state.
Thus, replacing the Korringa term $k_BT$ by a $d$-wave gapped term $k_BT$($k_BT/\Delta_{max})^2$
(2$\Delta_{max}=8k_BT_c$~\cite{Hasegawa}), 
one estimates $\left|JN_F\right|$=0.009 and then $\left|J\right|$=5.8 meV for Cu(2)
(if taking into account the effect of the antiferromagnetic spin correlation, one gets a smaller $\left|J\right|$). 
The estimated magnitude of $\left|J\right|$ is smaller than the in-plane exchange interaction $\left|J_{Ni-Ni}\right|$=11$\sim$31 meV of
La$_2$NiO$_{4+\delta}$~\cite{Yamada,Nakamura}, 
$\left|J_{Cu-Cu}\right|$=150 meV of YBa$_2$Cu$_3$O$_{6+\delta}$~\cite{Rossat}, or
a typical 4$s$-3$d$ exchange interaction $\left|J_{sd}\right|\sim$0.1 eV.  
Since $\left|\pi JN_FS\right|=0.03\ll$1, the decrease of $T_c$ due to the magnetic scattering can be
calculated within the lowest order Born approximation~\cite{AG}.
From $\Delta T_c=0.25\pi^2x_{plane}N_FJ^2S(S+1)/k_B$ with
$S$=1 in a $d_{x^2-y^2}$-wave superconductor~\cite{Maekawa}, the sole occupation at Cu(2) ($x_{plane}$=$x$/2=0.05) in the quenched sample
yields $\Delta T_c$=0.08 K at most. 
This value is smaller than the observed $\Delta T_c\sim$20 K in the quenched sample by a factor $\sim$250, 
suggesting that magnetic pair breaking is not the mechanism of suppression of $T_c$ in YBa$_2$Cu$_{3-x}$Ni$_x$O$_{7-\delta}$ 
(similar estimation for Zn doping is seen in Ref.~\cite{Walstedt}). 

Here, one may doubt the assumption of $\Gamma_L$=$\omega_N$ at the temperature of maximum 1/$\tau_1$. 
Instead of $\Gamma_L$=$\omega_N$, one may associate the maximum of 1/$\tau_1$ with a minimum behavior of $\Gamma_L(\gg \omega_N$) 
as a crossover from the high temperature $\Gamma_{SL}$ to the low temperature $\Gamma_{SS}$.
As mentioned above, $\Gamma_{SS}$ is proportional to $x_{plane}$, so that the crossover temperature must increase
with increasing $x_{plane}$. 
However, the temperature of maximum 1/$\tau_1$ is nearly the same in two samples.
Thus, this scenario is unlikely.    
In passing, if the Ni spin freezing temperature $T_M\sim$1.5 K ($x\sim$0.09) in Ref.~\cite{Tokunaga2}
is assigned to the maximum temperature $T=\omega_N/\alpha$, one obtains $\left|JN_F\right|$=1.9 and then $\left|J\right|$=1.3 eV.
This magnitude of $\left|J\right|$ is too large, although it can easily reproduce the observed $\Delta T_c$. 
In addition, the Ni spin fluctuation frequency $\Gamma_L$ increases up to $\sim$20 meV at $T_c$ due to the large $\left|J\right|$, 
so that the neutron scattering technique must probe the increase of magnetic response around 20 meV due to Ni spin. 
But, the existing data do not support the observation of Ni spin fluctuation~\cite{Sidis}. 
Thus, $T_M\sim\omega_N/\alpha$ is also unlikely.   

From $\Delta T_c=0.25\pi^2x_{plane}N_FJ^2S(S+1)/k_B$ with 0$<x_{plane}<$0.05 in a $d_{x^2-y^2}$-wave superconductor~\cite{Maekawa},
$\left|J\right|>$68 meV ($S$=1) or $\left|J\right|>$111 meV ($S$=1/2) is required to account for $\Delta T_c\sim$20 K
in the quenched sample (similar estimation for Zn doping is seen in Refs.~\cite{Mahajan,Monod}).
Then, $\left|\pi N_FJS\right|>$0.3 indicates moderately strong scattering.  
The expression of $\Delta T_c$ should be extended to include
both potential scattering and strong scattering. 

In considering the magnetic impurity effect in $s$-wave superconductors,
one can neglect nonmagnetic potential scattering which should accompany 
each magnetic impurity, because of Anderson's theorem, unless the localization effect takes place. 
However, the situation is completely different in the $d$-wave state. 
The potential scattering also affects $d$-wave superconductivity 
as well as the magnetic impurity scattering; thus one has to take into account both effects (see also ~\cite{Maekawa,Salkola,Kilian}). 
Let us treat magnetic impurities as classical spins ($S_z=\pm S$) 
and assume that they also have nonmagnetic potential scatterings described by $u$. 
Taking into account the impurity scatterings 
within a $t$-matrix approximation, Ohashi derived the theoretical expression of $T_c$ ($x_{plane}\ll$1) in a $d_{x^2-y^2}$-wave
superconductor~\cite{Ohashi},
\begin{equation}
\Delta T_c={x_{\rm plane} \over 4k_BN_F}
\bigl[
1-{1 \over 2}
(
{1 \over 1+(\gamma_n-\gamma_m)^2}+
{1 \over 1+(\gamma_n+\gamma_m)^2}
)
\bigr].
\label{eq.1}
\end{equation}
Here, $\gamma_n$ and $\gamma_m$, respectively, describe the depairing effects originating
from nonmagnetic and magnetic parts of the magnetic impurities:
\begin{eqnarray}
\left\{
\begin{array}{l}
\gamma_n=\pi N_Fu,\\
\gamma_m=\pi N_FJS.
\end{array}
\right.
\label{eq.2}
\end{eqnarray}
\par
Putting $\gamma_m=0$ in eq. (1), we obtain $\Delta T_c$ in the case of nonmagnetic impurity Zn,
being in agreement with the expression in Refs.~\cite{Borkowski,Fehrenbacher,Arberg,Kitaoka}. 
On the other hand, $\gamma_n=0$ gives $\Delta T_c$ in the case of magnetic impurity without nonmagnetic potential
scattering part. 
In the limit of $\gamma_m\rightarrow 0$ and $\gamma_n\rightarrow 0$, eq. (1) leads to the expression of $\Delta T_c$ in the lowest
order Born approximation, being in agreement with the expression in Ref.~\cite{Maekawa}. 
In addition, $\gamma_m\to\infty$ or $\gamma_n\to \infty$ corresponds to the unitarity limit~\cite{Hotta}; 
$\Delta T_c=x_{plane}/4k_BN_F\approx(1.9\times10^3$ K)$x_{plane}$.   
If cancellation of $\gamma_m-\gamma_n=0$ in the unitarity limit takes place for Ni, 
one obtains $\Delta T_c=x_{plane}/8k_BN_F$, which is two times smaller than that due to pure nonmagnetic scattering.
For YBa$_2$Cu$_{3-x}$Ni$_x$O$_{7-\delta}$, however, one should assume $\gamma_n\gg\gamma_m$.
In the moderately strong scattering (a finite $u$), 
since $\Delta T_c\propto x_{plane}N_F$ at $N_F\ll1/\pi u$ ($\gamma_n\ll1$) and $\Delta T_c\propto x_{plane}/N_F$ at $N_F\gg1/\pi
u$ ($\gamma_n\gg1$), then $\Delta T_c$ takes a maximum of 0.125$x_{plane}/k_BN_F$ at the optimal $N_F=1/\pi u$ ($\gamma_n=1$).
If $N_F$ of YBa$_2$Cu$_3$O$_{7-\delta}$ is optimal, that is $u$=1/$\pi N_F$=212 meV ($\gamma_n=1$),
using $\Delta T_c=(9.5\times10^2$ K)$x_{plane}$, we estimate $x_{plane}$=0.01 (as-synthesized) and 0.02 (quenched).
In the unitarity limit, eq. (1) does not explicitly depend on $u$ nor $J$. 
Then, using $\Delta T_c=(1.9\times10^3$ K)$x_{plane}$, we estimate the minimum values of $x_{plane}$=0.005 (as-synthesized) and
0.01 (quenched).
These values are so reasonable as to satisfy 0$<x_{plane}<$0.05. 
Thus, we suggest that the moderately strong, nonmagnetic scattering of Ni is a promising origin of $T_c$ suppression.
This is qualitatively consistent with
a report on a large $\left|u\right|(>\left| JS\right|$) of Ni in Bi$_2$Sr$_2$CaCu$_2$O$_{8+\delta}$~\cite{Hudson}.  

	In conclusion, the Cu NQR experiment demonstrated that both Cu(1) and Cu(2) nuclear spin-lattice relaxations
in YBa$_2$Cu$_{3-x}$Ni$_x$O$_{7-\delta}$ are affected by Ni doping,  
that is, the doped Ni impurities are substituted for both Cu(1)
and  Cu(2) sites. One of the reasons 
for the small decrease of $T_c$ by Ni doping in YBa$_2$Cu$_3$O$_{7-\delta}$ is partial substitution of the Ni impurities for the
chain site. In the light of the Ni-induced nuclear spin-lattice relaxation, the host Cu(2) spin fluctuation spectrum in
YBa$_2$Cu$_3$O$_{7-\delta}$ with optimal oxygen content is quite robust for Ni doping.   

	We would like to thank Professors M. Matsumura, H. Yamagata, and K. Mizuno 
for helpful discussion. This work was supported by the New Energy and Industrial 
Technology Development Organization (NEDO) as Collaborative Research and 
Development of Fundamental Technologies for Superconductivity Applications.

\begin{figure}
\epsfxsize=3.7in
\epsfbox{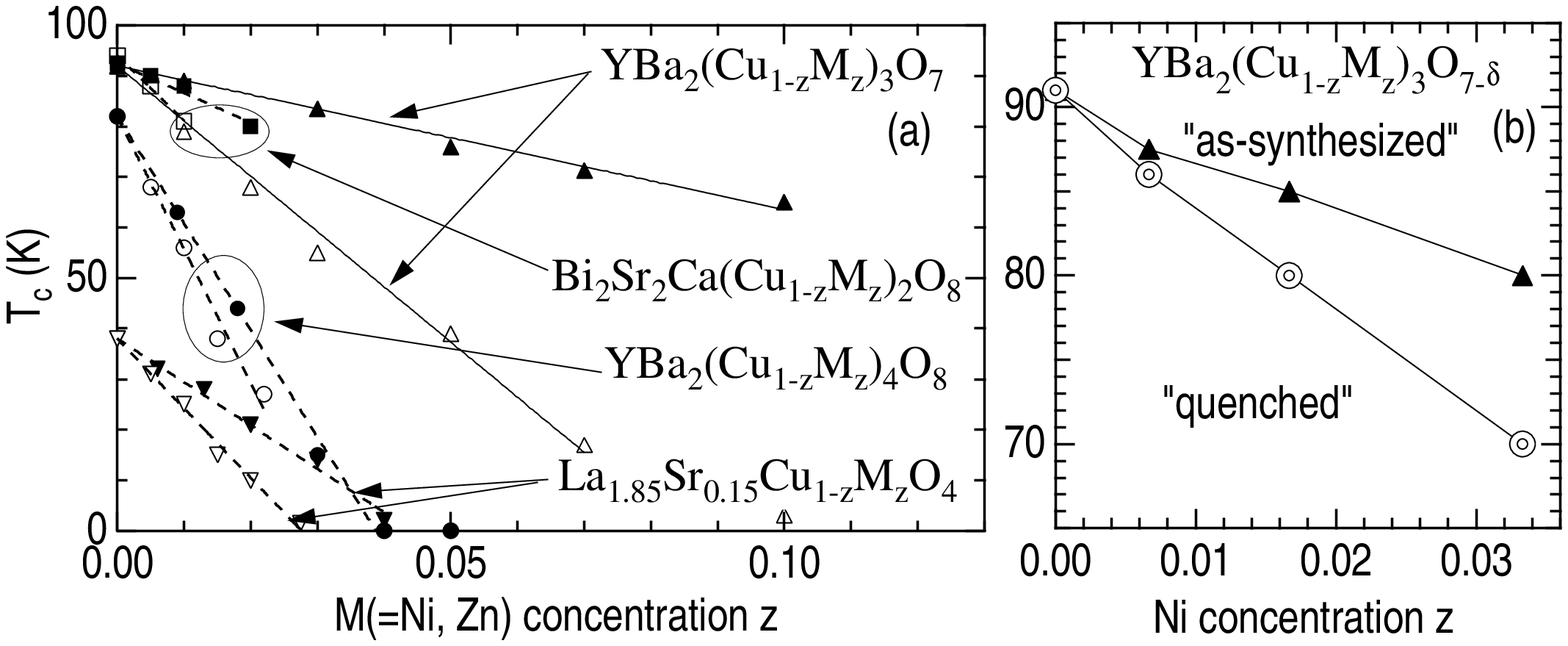}
\vspace{0.1cm}
\caption{
(a)Impurity doping dependence of $T_c$ as functions of the concentration $z$ for various high-$T_c$ superconductors
with M=Ni or Zn.
The solid (open) symbols are the Ni(Zn) doping dependence of $T_c$.
The data are adopted from [9] for YBa$_2$(Cu$_{1-z}$M$_z$)$_3$O$_7$ (Y123, upward triangles), [10] for
Bi$_2$Sr$_2$Ca(Cu$_{1-z}$M$_z$)$_2$O$_8$ (Bi2212, squares), [11,12] for YBa$_2$(Cu$_{1-z}$M$_z$)$_4$O$_8$ (Y124,
circles), and [1] for La$_{1.85}$Sr$_{0.15}$Cu$_{1-z}$M$_z$O$_4$ (LS214, downward triangles). 
The dashed and the solid lines are fit by $T_c$=$T_{c0}$-$m_Mz$ ($m_{M=Ni, Zn}$ is the fitting parameter) 
for the respective materials.  
The estimated ratio $m_{Ni}$/$m_{Zn}$ is about 0.26 for Y123, 0.46 for Bi2212, 0.80 for Y124,
and 0.62 for LS214. The decrease of $T_c$ by Ni doping for Y123 is smallest among these materials.
(b)Ni doping dependence of $T_c$ for "as-synthesized" or "quenched" Y123, adopted from [18]. 
The solid curves are guide for the eyes.
Note that the Ni concentration $x$ in the text is defined by $x$=3$z$ in YBa$_2$Cu$_{3-x}$M$_x$O$_{7-\delta}$
(7-$\delta$=6.92-6.95 [18]).
The decrease of $T_c$ by Ni doping is larger in "quenched" Y123 than in "as-synthesized" one. 
}
\label{Tcdata}
\end{figure}

\begin{figure}
\epsfxsize=3.5in
\epsfbox{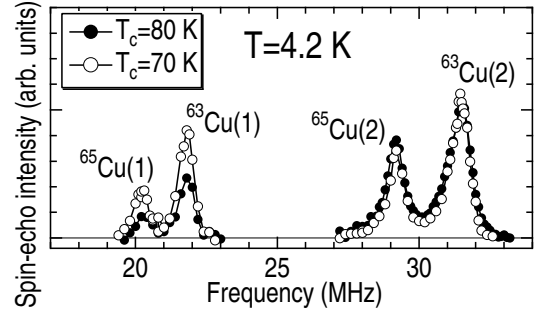}
\vspace{0.1cm}
\caption{
Zero-field frequency spectra of the chain-site and the plane-site 
$^{63, 65}$Cu NQR for YBa$_2$Cu$_{3-x}$Ni$_x$O$_{7-\delta}$ at $T$=4.2 K.
The line shapes are scaled after $T_2$ corrections have been made. 
}
\label{NQR}
\end{figure}
   
\begin{figure}
\epsfxsize=3.5in
\epsfbox{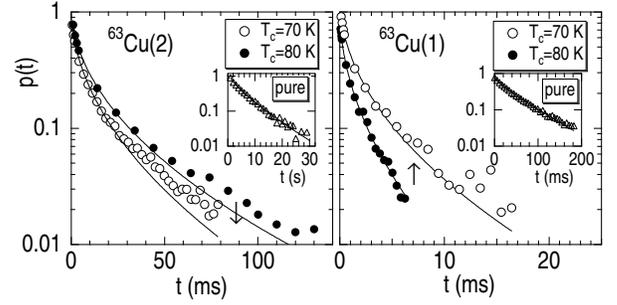}
\vspace{0.1cm}
\caption{
The Ni-doping effect on the $^{63}$Cu nuclear spin-echo recovery
curves $p(t)\equiv1-M(t)/M(\infty)$ at $T$=4.2 K. 
Inset figures show the recovery curves for the pure (Ni-free) sample at $T$=4.2 K. 
The solid curves are the least-squares
fitting results using the theoretical function including a stretched exponential function (see the text).
}
\label{Recovery}
\end{figure}

\begin{figure}
\epsfxsize=3.5in
\epsfbox{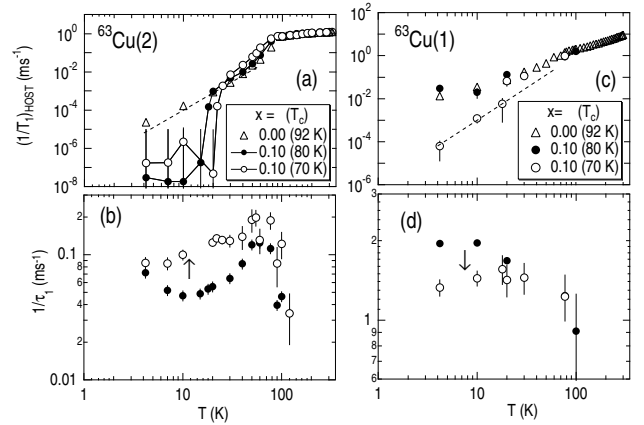}
\vspace{0.1cm}
\caption{
Log-log plots of $^{63}(1/T_1)_{HOST}$ (a) [(c)]
and $^{63}(1/\tau_1$) (b) [(d)] at $^{63}$Cu(2) [$^{63}$Cu(1)] 
as functions of temperature 
for the Ni-free and for the Ni-doped samples. 
The dashed lines are $T^3$ functions. 
The solid curves are guide for the eyes. 
}
\label{LogT1}
\end{figure}

\begin{figure}
\epsfxsize=3.7in
\epsfbox{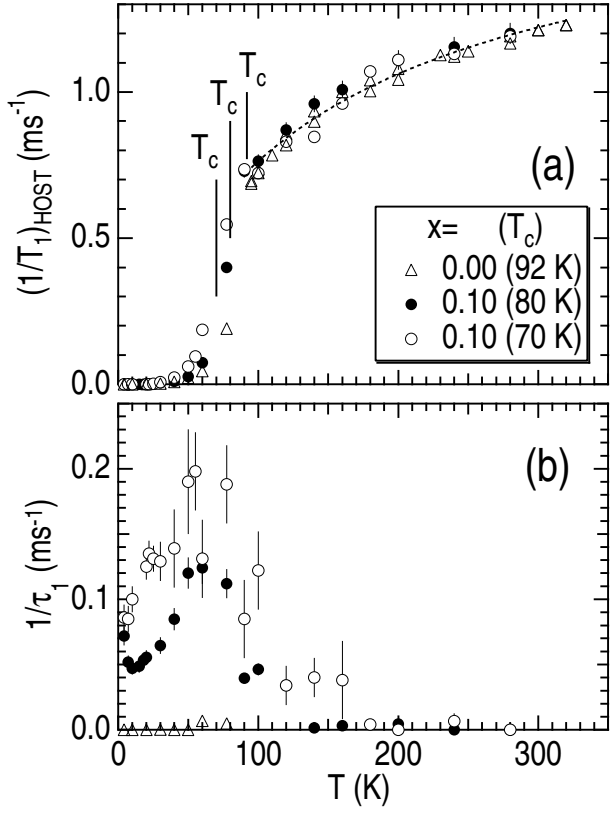}
\vspace{0.1cm}
\caption{
The Ni-doping effect on the temperature dependence of $^{63}(1/T_1)_{HOST}$ (a) 
and $^{63}(1/\tau_1$) (b) at $^{63}$Cu(2)  
in linear scale. The thick bars indicate the respective $T_c$'s. The dotted curve is the least-squares fitting result
by a function of $TC/(T+\Theta)$ ($C$=1.7 $ms^{-1}$ and $\Theta$=128 $K$) above $T_c$.
}
\label{Linear T1}
\end{figure}

\end{document}